# Luminescence anomaly of dipolar valley excitons in homobilayer semiconductor moiré superlattices

Hongyi Yu[1*], Wang Yao[2,3*]

[1] Guangdong Provincial Key Laboratory of Quantum Metrology and Sensing & School of Physics and Astronomy, Sun Yat-Sen University (Zhuhai Campus), Zhuhai 519082, China
[2] Department of Physics, The University of Hong Kong, Hong Kong, China
[3] HKU-UCAS Joint Institute of Theoretical and Computational Physics at Hong Kong, China
[*] Corresponding authors: yuhy33@mail.sysu.edu.cn (HY), wangyao@hku.hk (WY).

**Abstract:** In twisted homobilayer transition metal dichalcogenides, intra- and inter-layer valley excitons hybridize with the layer configurations spatially varying in the moiré. The ground state valley excitons are trapped at two high-symmetry points with opposite electric dipoles in a moiré supercell, forming a honeycomb superlattice of nearest-neighbor dipolar attraction. We find that the spatial texture of layer configuration results in a luminescence anomaly of the moiré trapped excitons, where a tiny displacement by interactions dramatically increases the brightness and changes polarization from circular to linear. At full filling, radiative recombination predominantly occurs at edges and vacancies of the exciton superlattice. The anomaly also manifests in the cascaded emission of small clusters, producing chains of polarization entangled photons. An interlayer bias can switch the superlattice into a single-orbital triangular lattice with repulsive interactions only, where the luminescence anomaly can be exploited to distinguish ordered states and domain boundaries at fractional filling.

Long-wavelength moiré patterns by van der Waals stacking of two-dimensional crystals has made possible the exploration into a new realm of physics. The local-to-local variation of the interlayer registry leads to the spatial modulation of the local band structure, creating an artificial superlattice with a period in the range from a few to a few tens nanometers. In twisted graphene moiré, interlayer coupling at magic angles turns the massless Dirac cones into flat mini-bands, where a plethora of correlation phenomena are observed [1-9]. Another exciting moiré platform is the semiconducting transition metal dichalcogenides (TMDs) which host massive Dirac fermions at a time-reversal pair of valleys located at the **K** and −**K** corners of the hexagonal Brillouin zone [10]. Moiré pattern in TMDs heterostructures introduces a triangular superlattice landscape for valley electrons [11], in which a variety of correlated insulating states have been observed at various filling factors [12-17].

Excitons formed by the Coulomb binding of valley electron and hole make possible a distinct many-body system in the TMD moiré superlattices with optical addressability and the bosonic statistics. The intralayer exciton ($X_{intra}$) has a large optical dipole with a fixed valley polarization selection rule, which underlies the versatile optical controls of valley dynamics [18-21]. In type-II heterobilayers, excitons also form in an interlayer configuration ($X_{inter}$), where the separation of constituent electron and hole into adjacent layers leads to a permanent electric dipole that turns on strong dipolar interactions and coupling to electric field [22-27]. The layer separation of the electron and hole leads to long recombination lifetime and valley lifetime of $X_{inter}$, favorable for valleytronic applications [28], as well as for the exploration of quantum many-body phenomena such as exciton Bose-Einstein condensation [29] and self-assembled crystal phase of the two-dimensional dipolar excitons [30].

In the moiré pattern, $X_{inter}$ experiences a strong superlattice potential with spatial variation in the order of ~100 meV, and features an optical dipole that spatially varies on the scale of the moiré period, determined by the local stacking registry [31-41]. The lowest energy $X_{inter}$ are then trapped at the potential minima, which can form an ordered array of quantum emitters. The radiative recombination of these $X_{inter}$ ground states results in sharp emission lines with spectral width of ~ 0.1 meV, as observed in MoSe$_2$/WSe$_2$ heterostructures under low excitation power and temperature [34-36]. Photon antibunching experiment indicates that these trapped $X_{inter}$ can serve as highly tunable single-photon emitters [36]. The repulsive dipolar interaction between the trapped $X_{inter}$ manifests as blue shifts of their resonances observed in the photoluminescence spectrum [25-27]. At the potential minima, the **$2\pi/3$**-rotational symmetry ($\hat{C}_3$) of the local atomic registries dictates circularly polarized optical selection rules of $X_{inter}$, with the emission polarization jointly determined by the valley, spin and stacking registry [31,42]. The polarization properties of the exciton emission (co- or cross-polarized with respect to the excitation at $X_{intra}$ resonance) can be used to identify the trapping location within a moiré supercell, which is switchable by an out-of-plane electric field [31]. While these properties of $X_{inter}$ in the heterostructure moiré are highly intriguing for novel optoelectronic applications, the very weak optical dipole has however placed an intrinsic limitation on their exploitation [39].

Here we present an exotic exciton system in twisted TMDs homobilayers. We show that because of the registry dependent interlayer coupling, the energies of intra- and inter-layer valley excitons cross each other as functions of position in the moiré, giving rise to a texture of spatially varying exciton hybridization. In a moiré supercell, the ground state valley exciton is trapped at two degenerate high-symmetry points with opposite out-of-plane electric dipole moments, forming a honeycomb superlattice with repulsive on-site and attractive nearest-neighbor interactions. The spatial texture of the $X_{intra}$-$X_{inter}$ hybridization leads to a luminescence anomaly of excitons in the traps. At equilibrium positions, they are as dark as $X_{inter}$, while a tiny lateral displacement in the order of Å dramatically increases the brightness, and the emission polarization is changed from circular to linear, with polarization angle reflecting the displacement direction. At full exciton filling of the honeycomb superlattice, radiative recombination predominantly occurs at edges and vacancies where excitons are displaced by the unbalanced dipolar interactions from neighboring sites. An interlayer bias can switch the excitonic superlattice to a single-orbital triangular one with repulsive interactions only. We show that the luminescence anomaly can serve as signatures for different correlated states and domain boundaries at fractional exciton filling. The anomaly also manifests in the cascaded emission of small clusters, producing chains of polarization entangled photons.

Consider the near 0° twisted homobilayers with moiré period $b$ much larger than the monolayer lattice constant, such that local regions can be approximated by commensurate bilayers with various R-stacking registries (Fig. 1a). For the conduction and valence band edge electrons at **K** valley, interlayer coupling has a sensitive dependence on the stacking registry and therefore becomes a periodic function of position **R** in the moiré [43,44]:

$$\widehat{H}_{CBM}(\mathbf{R}) = \begin{pmatrix} E_c(\mathbf{R}) & h_{cc'}(\mathbf{R}) \\ h_{cc'}^*(\mathbf{R}) & E_{c'}(\mathbf{R}) \end{pmatrix}, \qquad \widehat{H}_{VBM}(\mathbf{R}) = \begin{pmatrix} E_v(\mathbf{R}) & h_{vv'}(\mathbf{R}) \\ h_{vv'}^*(\mathbf{R}) & E_{v'}(\mathbf{R}) \end{pmatrix}.$$

Here $E_{c/v}(\mathbf{R})$ is the energy shift of the conduction/valence band edge in the upper layer due to the interlayer coupling, while $E_{c'/v'}(\mathbf{R})$ is the one in the lower layer. $h_{cc'/vv'}(\mathbf{R})$ is the interlayer hopping of the electron/hole. Their **R**-dependences underlie the spatially-varying layer hybridizations of the valley electrons and holes in the moiré of TMDs homobilayers [43,44]. The phenomena to be explored are more pronounced for smaller $b$ ($\lesssim$ **10** nm), where the lattice reconstruction effect is insignificant [45], not considered here.

We focus on the valley excitons formed by a pair of electron and hole at these band edges. The strong electron-hole Coulomb interaction leads to a binding energy of a few hundred meV, which is one order larger than the interlayer hopping ($|h| \sim 20$ meV). This allows one to start from the exciton basis obtained in the limit of vanishing interlayer hopping. The basis states in a given valley include the two X$_{\text{inter}}$, denoted as $|cv'\rangle_\mathbf{R}$ and $|c'v\rangle_\mathbf{R}$ which have opposite electric dipoles (Fig. 1b), and the two X$_{\text{intra}}$ confined in the two layers respectively, denoted as $|cv\rangle_\mathbf{R}$ and $|c'v'\rangle_\mathbf{R}$ (Fig. 1b). In this basis, the exciton Hamiltonian reads,

$$\widehat{H}_X(\mathbf{R}) = E_0 - \frac{\hbar^2 \nabla^2}{2M} + \widehat{V}(\mathbf{R}), \qquad (1)$$

$$\widehat{V}(\mathbf{R}) \equiv \begin{pmatrix} E_{cv'}(\mathbf{R}) + \Delta E_b & 0 & h_{vv'}(\mathbf{R}) & h_{cc'}(\mathbf{R}) \\ 0 & E_{c'v}(\mathbf{R}) + \Delta E_b & h_{cc'}^*(\mathbf{R}) & h_{vv'}^*(\mathbf{R}) \\ h_{vv'}^*(\mathbf{R}) & h_{cc'}(\mathbf{R}) & E_{cv}(\mathbf{R}) & 0 \\ h_{cc'}^*(\mathbf{R}) & h_{vv'}(\mathbf{R}) & 0 & E_{c'v'}(\mathbf{R}) \end{pmatrix}.$$

Here $E_0$ is the exciton energy in a pristine monolayer, and $M$ is the exciton effective mass. $\Delta E_b$ is the binding energy difference between X$_{\text{intra}}$ and X$_{\text{inter}}$. The electron/hole interlayer hopping causes the off-diagonal couplings between X$_{\text{intra}}$ and X$_{\text{inter}}$. $E_{ij}(\mathbf{R}) \equiv E_i(\mathbf{R}) - E_j(\mathbf{R})$ ($i = c/c'$, $j = v/v'$) accounts for the energy shift of X$_{\text{intra}}$ and X$_{\text{inter}}$ by the position dependent interlayer coupling in the moiré, which can be determined from fitting first-principles calculations [46],

$$\begin{aligned} E_c(\mathbf{R}) &= -\delta_c\big(f_+(\mathbf{R}) + f_-(\mathbf{R})\big) + \Delta_c\big(f_+(\mathbf{R}) - f_-(\mathbf{R})\big), \\ E_{c'}(\mathbf{R}) &= -\delta_c\big(f_+(\mathbf{R}) + f_-(\mathbf{R})\big) - \Delta_c\big(f_+(\mathbf{R}) - f_-(\mathbf{R})\big), \\ E_v(\mathbf{R}) &= -\delta_v\big(f_+(\mathbf{R}) + f_-(\mathbf{R})\big) + \Delta_v\big(f_+(\mathbf{R}) - f_-(\mathbf{R})\big), \\ E_{v'}(\mathbf{R}) &= -\delta_v\big(f_+(\mathbf{R}) + f_-(\mathbf{R})\big) - \Delta_v\big(f_+(\mathbf{R}) - f_-(\mathbf{R})\big). \end{aligned} \qquad (2)$$

$f_\pm(\mathbf{R}) \equiv \frac{1}{9}\left|e^{i\delta\mathbf{K}\cdot\mathbf{R}} + e^{i(\hat{C}_3\delta\mathbf{K}\cdot\mathbf{R}\pm\frac{2\pi}{3})} + e^{i(\hat{C}_3^2\delta\mathbf{K}\cdot\mathbf{R}\pm\frac{4\pi}{3})}\right|^2$ are the lowest harmonics that satisfy the rotational and translational symmetry, which turn out to be sufficient in accounting for the energy modulation in the moiré [46]. $\delta\mathbf{K} = \frac{4\pi}{3b}(\mathbf{0},\mathbf{1})$, $\hat{C}_3\delta\mathbf{K}$ and

$\hat{C}_3^2 \delta \mathbf{K}$ are the three corners of the mini Brillouin zone related by $2\pi/3$-rotation ($\hat{C}_3$). For the example of bilayer MoTe$_2$ which has a direct bandgap, DFT calculations give $\Delta_c \approx 31$ meV, $\Delta_v \approx 42$ meV, $\delta_c \approx 2$ meV, $\delta_v \approx 0.5$ meV [44].

The energies of X$_{\text{intra}}$ and X$_{\text{inter}}$ (i.e., diagonal terms of $\hat{V}(\mathbf{R})$ in Eq. (1)) as functions of $\mathbf{R}$ obtained from Eq. (2) are shown in Fig. 1c, where the three high-symmetry locations in a supercell have the coordinates $A = \mathbf{0}$, $B = \left(\frac{b}{\sqrt{3}}, 0\right)$ and $C = \left(\frac{2b}{\sqrt{3}}, 0\right)$, respectively. The energy minima are located at $B$ and $C$, where X$_{\text{inter}}$ has the energy: $E_0 - (\delta_c - \delta_v) - (\Delta_c + \Delta_v) + \Delta E_b$, and X$_{\text{intra}}$ has the energy: $E_0 - (\delta_c - \delta_v) - (\Delta_v - \Delta_c)$. If $\Delta E_b < 2\Delta_c$, then the energies of X$_{\text{intra}}$ and X$_{\text{inter}}$ cross each other in the moiré, and the lowest energy branch is of the mixed nature, being X$_{\text{inter}}$ near $B$ and $C$, and majorly X$_{\text{intra}}$ elsewhere. Numerical calculations of $\Delta E_b$ fall in the range of 20-90 meV [40,47-49], while experiments in various hetero- and homobilayer TMDs report $\Delta E_b$ values in the range of 10-60 meV [48-52]. The plot in Fig. 1 has used $\Delta E_b = 40$ meV as an example of the $\Delta E_b < 2\Delta_c$ regime which we will discuss first. In the opposite limit of $2\Delta_c < \Delta E_b$, the lowest energy branch is X$_{\text{intra}}$ at zero interlayer bias, while a finite interlayer bias can also introduce the energy crossing of X$_{\text{intra}}$ and X$_{\text{inter}}$ in the moiré, which will be discussed next (c.f. Fig. 4a).

The interlayer hopping of electron and hole then lead to hybridization of X$_{\text{intra}}$ and X$_{\text{inter}}$, and anti-crossing of the exciton branches [38,53-55]. The position-dependence of the interlayer hopping can be similarly obtained [43,44]

$$h_{cc'/vv'}(\mathbf{R}) = -t_{c/v}(\mathbf{R})\left(e^{i\delta \mathbf{K} \cdot \mathbf{R}} + e^{i\hat{C}_3 \delta \mathbf{K} \cdot \mathbf{R}} + e^{i\hat{C}_3^2 \delta \mathbf{K} \cdot \mathbf{R}}\right). \quad (3)$$

$t_{c/v}$ are real numbers in the order of a few meV (see Supplementary Section I).

$\hat{H}_X(\mathbf{R})$ can then be solved by first diagonalizing its $\hat{V}(\mathbf{R})$ part, which leads to four decoupled branches ($|n = 1,2,3,4\rangle$) of layer-hybridized excitons with both X$_{\text{intra}}$ and X$_{\text{inter}}$ components. The eigenvalues, $V_1(\mathbf{R}), V_2(\mathbf{R}), V_3(\mathbf{R}), V_4(\mathbf{R})$, correspond to the moiré superlattice potentials for these decoupled branches, which, together with the kinetic energy $-\frac{\hbar^2 \nabla^2}{2M}$, lead to exciton minibands. The moiré potentials for the four branches are shown as solid thick curves in Fig. 1d, where the colormap indicates strength of the optical dipole, determined by the X$_{\text{intra}}$ weighting (c.f. Fig. 1e). The electric dipole determined by the X$_{\text{inter}}$ weighting is shown in Fig. 1f. At $A$, the local stacking has the out-of-plane mirror symmetry, where the exciton branches are either layer symmetric with large optical dipole, or antisymmetric which is dark. At $B$ and $C$ both $h_{cc'}$ and $h_{vv'}$ vanish, so the four branches at these two locations reduce to either X$_{\text{intra}}$ or X$_{\text{inter}}$, without layer hybridization.

Below we focus on the lowest energy branch $|1\rangle$, where the moiré potential $V_1(\mathbf{R})$ has two degenerate minima at $B$ and $C$ with opposite electric dipoles, corresponding to the interlayer configurations $|cv'\rangle_B$ and $|c'v\rangle_C$, respectively [56]. Away from these potential minima, branch $|1\rangle$ starts to pick up X$_{\text{intra}}$ component, because of the finite interlayer hopping which takes the chiral forms $h_{cc'/vv'}(B + \delta \mathbf{L}) = J_{c/v} \frac{\delta L}{b} e^{-i\theta} + O(\delta L^2)$, $h_{cc'/vv'}(C + \delta \mathbf{L}) = -J_{c/v} \frac{\delta L}{b} e^{-i\theta} + O(\delta L^2)$, with $\delta \mathbf{L} = \delta L(\cos\theta, \sin\theta)$ the small displacement vector. $J_c \sim 25$ meV and $J_v \sim 110$ meV are determined by the parameters $t_c$ and $t_v$ in Eq. (3). The perturbative expansion of $|1\rangle$ in the neighborhood of $B/C$ reads

$$|1\rangle_{B+\delta \mathbf{L}} \approx |cv'\rangle_{B+\delta \mathbf{L}} - \frac{J_v}{2\Delta_v - \Delta E_b}\frac{\delta L e^{i\theta}}{b}|cv\rangle_{B+\delta \mathbf{L}} - \frac{J_c}{2\Delta_c - \Delta E_b}\frac{\delta L e^{i\theta}}{b}|c'v'\rangle_{B+\delta \mathbf{L}}, \quad (4a)$$

$$|1\rangle_{C+\delta \mathbf{L}} \approx |c'v\rangle_{C+\delta \mathbf{L}} + \frac{J_c}{2\Delta_c - \Delta E_b}\frac{\delta L e^{i\theta}}{b}|cv\rangle_{C+\delta \mathbf{L}} + \frac{J_v}{2\Delta_v - \Delta E_b}\frac{\delta L e^{i\theta}}{b}|c'v'\rangle_{C+\delta \mathbf{L}}. \quad (4b)$$

It is important to note that $X_{intra}$ and $X_{inter}$ components in the above expressions feature distinct valley polarization selection rules. Consider excitons in **K** valley, $X_{intra}$ ($|cv\rangle_R$ and $|c'v'\rangle_R$) always emit $\sigma+$ circularly polarized photons [10], while the $X_{inter}$'s emission polarization is location and layer-configuration dependent [31,42]. The emission polarizations are $\sigma-$ and $z$ ($z$ and $\sigma-$) for $|cv'\rangle_B$ and $|c'v\rangle_B$ ($|cv'\rangle_C$ and $|c'v\rangle_C$), respectively, as indicated in Fig. 1d. The $X_{inter}$ components involved in Eq. (4a) and (4b) both emit $\sigma-$ photons.

Near $B/C$, the moiré potential $V_1(\mathbf{R})$ has a parabolic form $\frac{1}{2}M\omega^2\delta L^2$ with a harmonic oscillator frequency $\omega \propto b^{-1}$ ($\hbar\omega \sim 10$ meV when $b = 8$ nm, in agreement with the results in Ref. [57]). The ground state valley exciton is then a gaussian wavepacket trapped at $B/C$, with a half-width $w = \sqrt{\frac{\hbar}{M\omega}}$. For $b = 8$ nm, $w \sim 1.4$ nm, so hopping between the trapping sites becomes negligible. We examine the ground state exciton wavepacket: $|W\rangle_{\mathbf{R}_c} \equiv \frac{1}{w\sqrt{\pi}}\int d\mathbf{R} e^{-\frac{(\mathbf{R}-\mathbf{R}_c)^2}{2w^2}}|1\rangle_\mathbf{R}$, and analyze how its optical properties vary upon displacement of the potential minima by an in-plane force. The optical dipole of such a wavepacket is determined by its central location $\mathbf{R}_c$ (c.f. Supplementary Section II),

$$\mathbf{D}_B \equiv \langle \mathbf{0}|\widehat{\mathbf{D}}|W\rangle_{B+\delta \mathbf{L}} \approx D_I \mathbf{e}_-^* - \eta\frac{\delta L}{b}e^{i\theta}D_0\mathbf{e}_+^*, \quad (5)$$

$$\mathbf{D}_C \equiv \langle \mathbf{0}|\widehat{\mathbf{D}}|W\rangle_{C+\delta \mathbf{L}} \approx D_I \mathbf{e}_-^* + \eta\frac{\delta L}{b}e^{i\theta}D_0\mathbf{e}_+^*.$$

Here $\mathbf{e}_\pm^* \equiv \left(\frac{\mathbf{x}\pm i\mathbf{y}}{\sqrt{2}}\right)^*$ is the unit vector for the $\sigma\pm$ polarization, $D_0$ ($D_I$) the dipole strength of an $X_{intra}$ ($X_{inter}$) wavepacket of half-width $w$. It is interesting to note that $|W\rangle_{\mathbf{R}_c}$ centered right at the $B/C$ point already has a non-negligible $X_{intra}$ fraction (see Fig. 1e), which, however, does not contribute to the dipole because of the destructive interference due to the $\delta L e^{i\theta}$ factor (also see Appendix A for a symmetry analysis). The contribution of the $X_{intra}$ fraction to the optical dipole is linearly proportional to the displacement of the wavepacket center from $B/C$, with the factor $\eta \equiv \frac{J_v}{2\Delta_v - \Delta E_b} + \frac{J_c}{2\Delta_c - \Delta E_b} \sim 3.5$.

Because of the opposite polarization of the $X_{intra}$ and $X_{inter}$ contributions, the photon emission of the displaced wavepacket is elliptically polarized. Fig. 2a shows the calculated degree of circular-polarization $P_{circ}$ as a function of central location in a moiré supercell, for a wavepacket in branch $|1\rangle$. $P_{circ}$ is rather close to $+1$ in most of the supercell except the two small regions around $B$ and $C$, where it quickly decreases to $-1$ at $B/C$. Such a behavior is due to the large ratio of $|D_0/D_I|$. The plot used $|D_0/D_I| = 20$ from first-principles result [31,42], consistent with experiments [58,59]. Thus a tiny displacement $\delta L$ of the wavepacket center can result in a substantial change to the optical dipole in Eq. (5).

Fig. 2b is an enlarged view of the polarization patterns around $B/C$. As can be seen, in contrast to the $\sigma-$ polarization at $B/C$, wavepacket centers located on a ring centered at $B/C$ exhibit 100% linearly polarized emissions. The ring radius can be as small as ~**0.1 nm** for a typical moiré period of $b = $ **8 nm**. The degree of linear polarization and the angle of its major axis $\theta_{\text{Linear}}$ are obtained from Eq. (5):

$$P_{\text{Linear}} = \frac{2\eta|D_I/D_0|}{|D_I/D_0|^2 + |\eta\,\delta L/b|^2}\frac{\delta L}{b}, \qquad (6)$$
$$\theta_{\text{Linear}} = -\left(\frac{1}{2}\theta + \frac{\pi}{4}\right) \pm \frac{\pi}{4}.$$

Here $+/-$ is for $B/C$. The polarization angle $\theta_{\text{Linear}}$ has a one-to-one correspondence with the displacement direction $\theta$ (Fig. 2b). $P_{\text{Linear}}$ as a function of $\delta L/b$ for three different values of $\Delta E_b$ is shown in Fig. 2c. Besides the linear polarization, the magnitude of the optical dipole, given by $|\mathbf{D}_{B,C}| = \sqrt{|D_I|^2 + \left|\eta\frac{\delta L}{b}D_0\right|^2}$, also increases rapidly with $\delta L$ as indicated in Fig. 2d. The oscillator strength of the exciton, proportional to its radiative recombination rate, is proportional to the modular square of the optical dipole. We expect our perturbative results of Eq. (4-6) to hold up to $\frac{\delta L}{b} \sim$ **0.1**, which corresponds to a ~50-fold increase for $|\mathbf{D}_{B/C}|^2$. Thus upon a tiny displacement of the exciton wavepacket from $B/C$, there can be orders of magnitude increase in its radiative recombination rate.

The electric dipoles of the trapped excitons lead to strong repulsive onsite, and attractive nearest-neighbor interactions. Consider a nearest-neighbor $BC$ pair of trapped excitons, their attractive dipolar interaction can be approximated by $-V_D = -\frac{e^2 d^2}{4\pi\epsilon_0}\frac{1}{(b/\sqrt{3})^3}$ in a suspended bilayer ($\sim -$ **10** meV for $b = $ **8 nm**), with $\epsilon_0$ the vacuum permittivity and $d \approx $ **0.7 nm** the interlayer distance. Taking into account the small but finite Bohr radius $a_B$, $V_D$ will get larger by a small percentage $\propto \frac{a_B^2}{b^2}$ as compared to the form above (see Appendix B for a more detailed discussion about the exciton-exciton interaction). The total energy of the pair of excitons is minimized through a symmetric displacement towards each other (c.f. Fig. 3a), with a magnitude $\delta L_0 = \frac{e^2}{4\pi\epsilon_0}\frac{27}{\omega}\left(\frac{d}{b}\right)^2$. $\delta L_0$ as a function of $b$ for three different $\Delta E_b$ values are shown in Fig. 3b. The displacement is small in the order of 0.1 nm, but can already lead to a substantial change in the emission polarization which can become 100% linear, and a dramatic increase in the emission rate by up to two orders of magnitude (c.f. Fig. 3b inset).

We note that this rapid change of optical properties is correlated with the change in the exciton trapping potential (due to the occupation of its near neighbors), in which the exciton energy is discretized. And the sharp contrast of polarization and oscillator strength is between the excitonic ground states in the displaced and undisplaced moiré potential traps. Upon the change of the potential trap (e.g., due to the annihilation of a neighboring exciton), the original exciton ground state wavepacket is no longer an eigenstate of the new potential trap. The emission properties are then determined by the competition between the radiative decay and the relaxation to the new ground state. As the latter relaxation time scale is much faster than the radiative decay, the radiative emission will predominately come from the ground state of the new potential trap (see Appendix C).

Below we first give a summary of the selection rule and emission polarization of the trapped excitons in the twisted homobilayer moiré. For an exciton wavepacket centered right at $B/C$, the combination of $\hat{C}_3$ and time reversal symmetries requires $\mathbf{K}$ and $-\mathbf{K}$ valleys to emit photons with $\sigma-$ and $\sigma+$ circular polarizations, respectively. Thus the photon emission from valley unpolarized excitons exhibits neither circular nor linear polarization. Time reversal symmetry breaking can come from circularly polarized pumping or applying an out-of-plane magnetic field, which introduces valley polarized exciton population that emits with circular polarization.

A tiny displacement from $B/C$ breaks the $\hat{C}_3$-symmetry of the wavepacket, and the corresponding photon emission becomes elliptically polarized for either valley. The $\mathbf{K}$ and $-\mathbf{K}$ valleys are still related by a time reversal, which dictates that their emission polarizations exhibit opposite helicities but the same linear polarization, with the linear polarization angle determined by the displacement direction (c.f. Fig. 2b). A 360°-anticlockwise rotation of the displaced wavepacket around $B/C$ results in a 360°-clockwise rotation of the linear polarization angle. It is important to note that, upon the same displacement vector, exciton wavepackets near $B$ and $C$ exhibit orthogonal linear polarization directions, which comes from the opposite signs for $h_{cc'lvv'}(B + \delta\mathbf{L})$ and $h_{cc'lvv'}(C + \delta\mathbf{L})$, as a consequence of spatial symmetry of the homobilayer moiré supercell.

This leads to the following emission patterns for a pair of trapped excitons taking into account their dipolar interaction:

(1) A nearest-neighbor pair occupying a $B$ and $C$ site respectively. With the opposite electric dipoles at the two sites, the interaction is attractive and tends to displace the two excitons in opposite directions towards each other. The resultant photon emission then has the same linear polarization angle at $B$ and $C$, while the helicity depends on both the valley index and displacement magnitude. The linear polarization angle can be used to track the spatial orientation of the nearest-neighbor pair, while the helicity can provide information on the site-specific valley configurations.

(2) A next-nearest-neighbor pair occupying two $B$ (or $C$) sites. The repulsive dipolar interaction displaces the two excitons in opposite directions away from each other. The photon emissions from the two displaced excitons carry orthogonal linear polarizations, and helicity determined by the valley configuration and displacement magnitude.

Next we consider how the dramatic change of optical properties by the tiny displacement of these dipolar excitons can manifest in the light emission of their many-body states. In a many-body configuration, each trapped exciton experiences the overall dipolar force from others, and can be displaced depending on the occupation of its neighboring sites. The leading order displacement are determined by the nearest-neighbors. When the three nearest-neighbors are all occupied or unoccupied, the force is balanced and the exciton is centered at *B*/*C* with zero displacement, which is essentially dark. The other six configurations shown in Fig. 3c all correspond to unbalanced forces. The center exciton at equilibrium is then displaced by $\delta L_0$ along six possible directions with a substantially increased brightness, and the angle of the linear polarization distinguishes the six cases. Taking $b = 5$ **nm** as an example, we estimate the trapped exciton has a radiative lifetime of 800 ns at a balanced site, and 10 ns at an unbalanced site [31]. On the other hand, the phonon-assisted equilibration timescale for $X_{\text{inter}}$ is estimated to be in the order of picosecond [47] (also see Appendix C for a detailed analysis on the relaxation process of trapped excitons), orders faster than the radiative lifetime.

The recombination of a trapped exciton breaks the dipolar force balance of its nearest-neighbors, which go to the new equilibrium in a time scale significantly shorter than the radiative lifetime (see Appendix C for details). The radiative emission of an exciton thus changes dramatically the emission timescale and polarizations of its nearest-neighbors, a new form of optical nonlinearity which leads to unique luminescence phenomena. At full filling of the lattice, unbalanced sites appear only at the edges or vacancy defects (c.f. Fig. 3d). The radiative annihilation of the excitons will predominantly occur at the edges and vacancies. Under partial filling, the attractive nearest-neighbor interaction leads to clusterization. Different recombination pathways of a small exciton cluster are characterized by the energies and linear polarizations of the cascaded photon emission, resulting in a chain of entangled photons. Fig. 3e illustrates two sets of recombination pathways for a lowest energy 4-site cluster, each can generate a polarization-entangled photon pair. Other possible recombination pathways (not shown) also exist, which can generate entangled photons with frequencies different from those in Fig. 3e. The atomically thin geometry of the bilayer and the dipolar nature of the exciton further point to various possibilities to isolate a small cluster of excitons for entangled photon generation, e.g. through using a biased metal tip or local gate [60], or creating strained local regions [26,61], or engineering the surrounding dielectric environment [62] (see Appendix D).

The degeneracy of *B* and *C* sites can be lifted by a finite interlayer bias, which lowers (increases) the energy of the exciton trapped at *B* (*C*) by $V_{\text{bias}}$. Below we focus on a large enough $V_{\text{bias}}$, such that only the *B* sites will be occupied which then form a triangular lattice with repulsive dipolar interactions only. In Fig. 4a, we show such a case with $\Delta E_b = $ **80 meV** $ > 2\Delta_c$ and $V_{\text{bias}} = $ **50** meV. In the lowest energy branch, exciton is dark at *B*, and the brightness increases fast with the displacement from this energy minimum. The emission polarization also changes dramatically under a tiny displacement from *B* (see Fig. 4b). Fig. 4c shows the equilibrium emission properties for a nearest-neighbor pair of trapped excitons, which are displaced away from each other by the dipolar repulsion. The two excitons emit with orthogonal linear polarizations, with enhanced emission rate. The equilibrium displacement magnitude

$\delta L_0$ is shown in Fig. 4d under various moiré of period $b$ and two different $V_{\text{bias}}$ values. $\delta L_0$ can be enhanced by tuning $V_{\text{bias}}$ to near $\Delta E_b - 2\Delta_c$.

For an unbalanced site in the triangular lattice, the different occupations of its six nearest-neighbor sites result in three different displacement magnitudes ($\delta L_0$, $\sqrt{3}\delta L_0$ and $2\delta L_0$). The degree of the linear polarization and the optical dipole strength under the displacement magnitude $\delta L_0$ ($2\delta L_0$) are shown as solid (dashed) curve in Fig. 4d inset. Fig. 4e shows the displacements and the correlated linear polarizations of unbalanced sites at the edges and vacancies of a full filling exciton lattice, where exciton radiative recombination is much faster than the balanced site inside the bulk.

Such a triangular exciton superlattice with the repulsive interactions can host the various ordered states under fractional fillings, as those observed for electrons and holes in $WS_2/WSe_2$ heterobilayer moiré superlattices [13,15-17]. Remarkably, the luminescence anomaly of the unbalanced sites can be exploited as a fingerprint for certain configurations. In Fig. 4f we give the examples of two different stripe phases under filling factor 1/2. In the upper panel of Fig. 4f, all the occupied sites are unbalanced with alternating displacements $\pm\delta L_0$ along $y$ direction. In such a state, all trapped excitons have enhanced brightness, and emit with two orthogonal linear polarizations. The lower panel shows another stripe phase, where all trapped excitons are at balanced sites except at the domain boundary. Excitons are thus brightened only at the domain boundaries, and the linear polarization gives the boundary orientation.

Finally, we note that the luminescence anomaly is the property of the **K** valley excitons. The above quantitative analysis uses homobilayer $MoTe_2$ as the example, which has a direct bandgap with conduction and valence band edges both located at **K** (as evidenced by experiments in [63-65]). This could be a highly promising system to exploit the luminescence anomaly. $WSe_2$ bilayer moiré induced by heterostrain can be another candidate system. Pristine $WSe_2$ bilayer has an indirect band gap (**K** to **Q**), but the energy is not far from the direct **K** to **K** gap. Experiments have shown that, under ~ 1% uniaxial tensile strain, the band gap of bilayer $WSe_2$ becomes the direct **K** to **K** one [66,67]. Having the bilayer subject to heterostrain (different strain magnitude in the two layers, e.g. when strain is applied through substrate) can be an alternative way to create homobilayer moiré superlattices other than the twisting [68]. In such case, we note that the lowest energy excitons trapped at the ***B/C*** sites in R-type $WSe_2$ homobilayer moiré will be the spin-triplet at **K** valleys, which, at the unbalanced sites, are subject to similar displacements due to the dipolar interactions as described in Fig. 3a-c. Their emission can be the phonon assisted process via the spin-singlet exciton discussed here [69,70], or through hybridization with the spin-singlet in an in-plane magnetic field [71,72], so the sharply contrasted emission dipoles of the singlet at balanced and unbalanced sites will also manifest in the luminescence.

The relative orders of the energy scales for exciton in the moiré trap are underlying the perturbative treatment of the layer hybridization (Eq. (4)) and exciton relaxation (see Appendix C), and allow us to focus only on the ground state in the trap for the radiative emission. The radiative recombination rate ($\leq 0.1$ ns$^{-1}$) is several orders of magnitude slower than the rate for the exciton to relax to the ground state in the trap (~ ps$^{-1}$, see Supplementary Section III). The latter rate is also much smaller than the energy detuning and off-diagonal coupling between $X_{inter}$ and $X_{intra}$, and the quantization energy in the trapping potential $\hbar\omega$ (**~10** meV for $b = $ **8** nm). It is worth noting that the detuning of $X_{inter}$ and $X_{intra}$ at the ***B/C*** trap is controllable by an interlayer bias and can also be affected by the dielectric environment (through their distinct binding energies). When $X_{inter}$ and $X_{intra}$ are brought closer in energy such that their detuning is no longer significantly larger than the phonon and photon induced broadening, corrections beyond the perturbative treatment becomes non-negligible. In such case, the luminescence phenomenon can be a more complex manifestation of the interplay between the interlayer hybridization, relaxation and dephasing by phonon, and radiative emission processes.

## Acknowledgements

WY acknowledges support by the National Key R&D Program of China (2020YFA0309600), and the Croucher Foundation (Croucher Senior Research Fellowship). HY acknowledges support by Guangdong project No. 2019QN01X061.

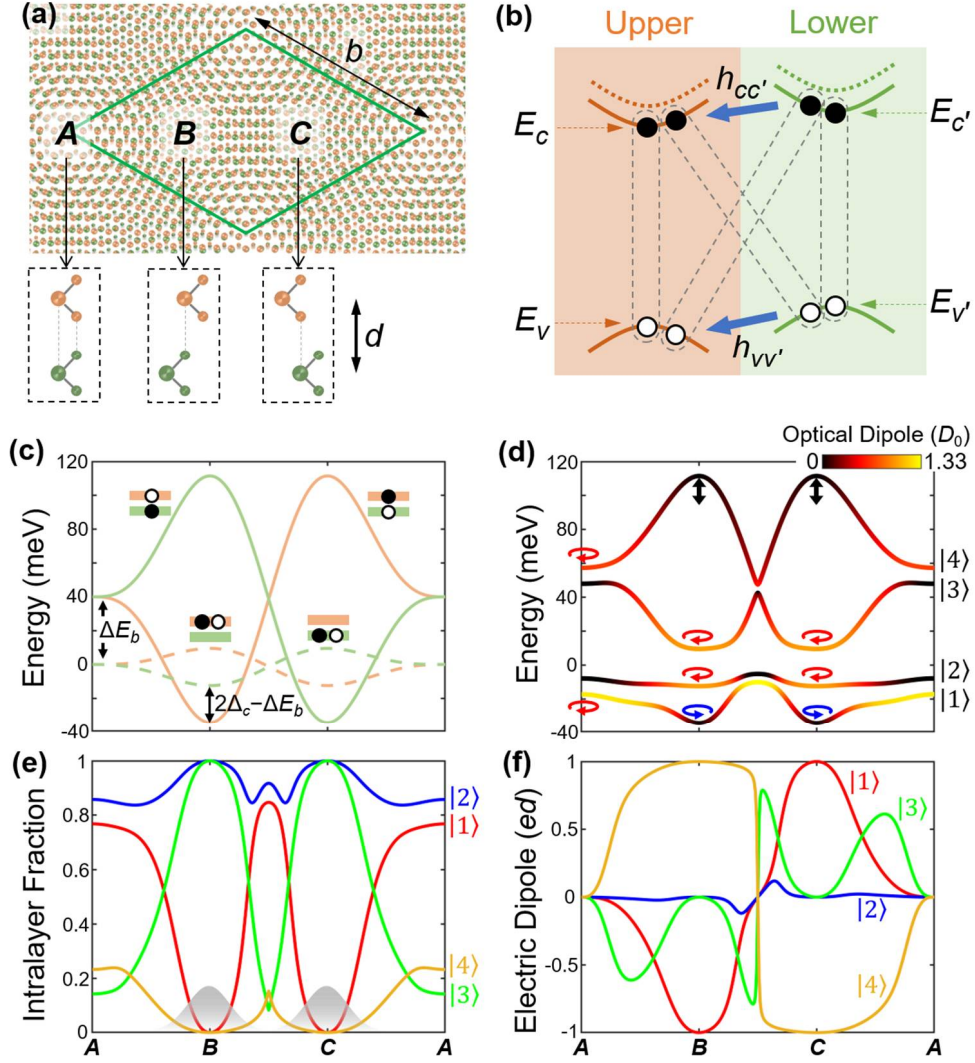

**Fig. 1 Hybridization of inter- and intra-layer excitons in a homobilayer moiré.** (a) Schematic illustration of a near R-type twisted homobilayer moiré. The green diamond is a supercell, which has three high-symmetry local regions **A**, **B** and **C** with different atomic registries. (b) Excitons of intralayer ($X_{intra}$) and interlayer ($X_{inter}$) configurations. At a general location in the moiré, the interlayer coupling manifests as the off-diagonal hopping of electron and hole ($h_{cc'}$ and $h_{vv'}$), and the diagonal energy shifts of the band edges in the two layers which results in a modest type-II band offset $\sim O(10)$ meV. (c) Solid (dashed) lines indicate the energies of $X_{inter}$ ($X_{intra}$), along the long diagonal of the supercell in (a). (d) Moiré potentials for the four decoupled exciton branches, each of which is a hybridization of $X_{inter}$ and $X_{intra}$ by the electron and hole interlayer hopping (c.f. main text). The optical dipole strength is color-coded in units of $D_0$, the optical dipole of $X_{intra}$. (e) Fraction of $X_{intra}$ in each branch. The shaded wavepackets indicate excitons trapped at **B** and **C** in the lowest energy branch $|1\rangle$. (f) The electric dipole of each branch in units of $ed$.

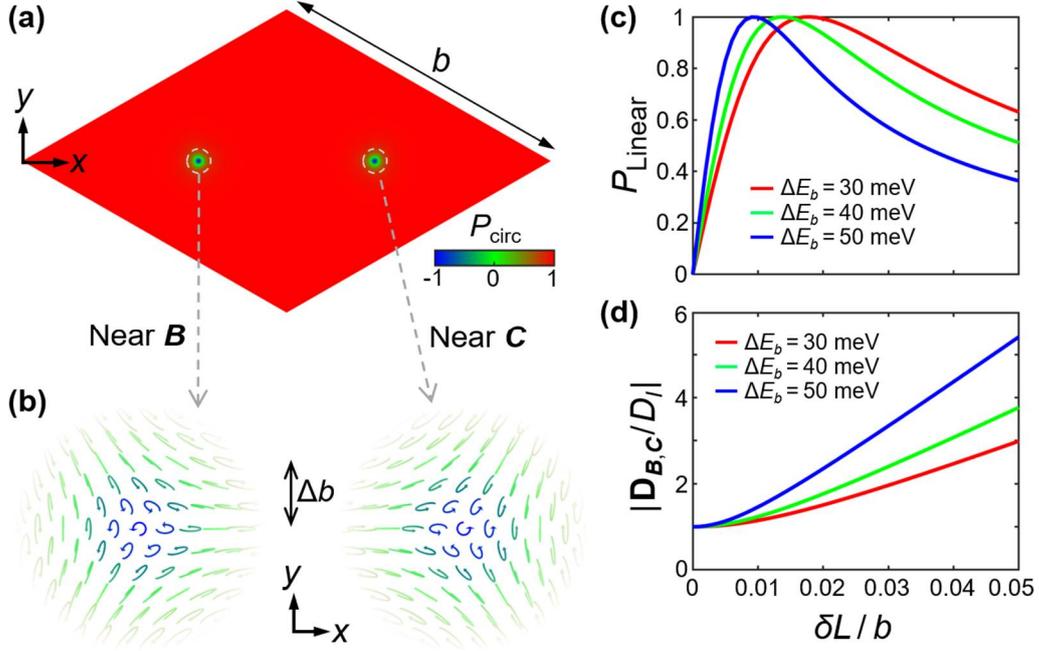

**Fig. 2 Luminescence anomaly in the moiré.** (a) Degree of circular polarization $P_{\text{circ}}$ of the photon emission by the lowest energy branch $|\mathbf{1}\rangle$, as a function of the wavepacket central position in a moiré supercell. $P_{\text{circ}} = 1$ over nearly the entire supercell, except the small neighborhood of **B** and **C**, where the rapid variation of $P_{\text{circ}}$ is due to the hybridization of $X_{\text{inter}}$ and $X_{\text{intra}}$. (b) The enlarged view of the emission polarization patterns near **B** and **C**, over a length scale of $\Delta b \sim 0.01b$. (c) Degree of linear-polarization $P_{\text{linear}}$ when an exciton wavepacket trapped at **B**/**C** has a tiny displacement of $\delta L$. (d) The optical dipole strength (in unit of $|D_I|$, c.f. Eq. (5)) also changes remarkably by $\delta L$.

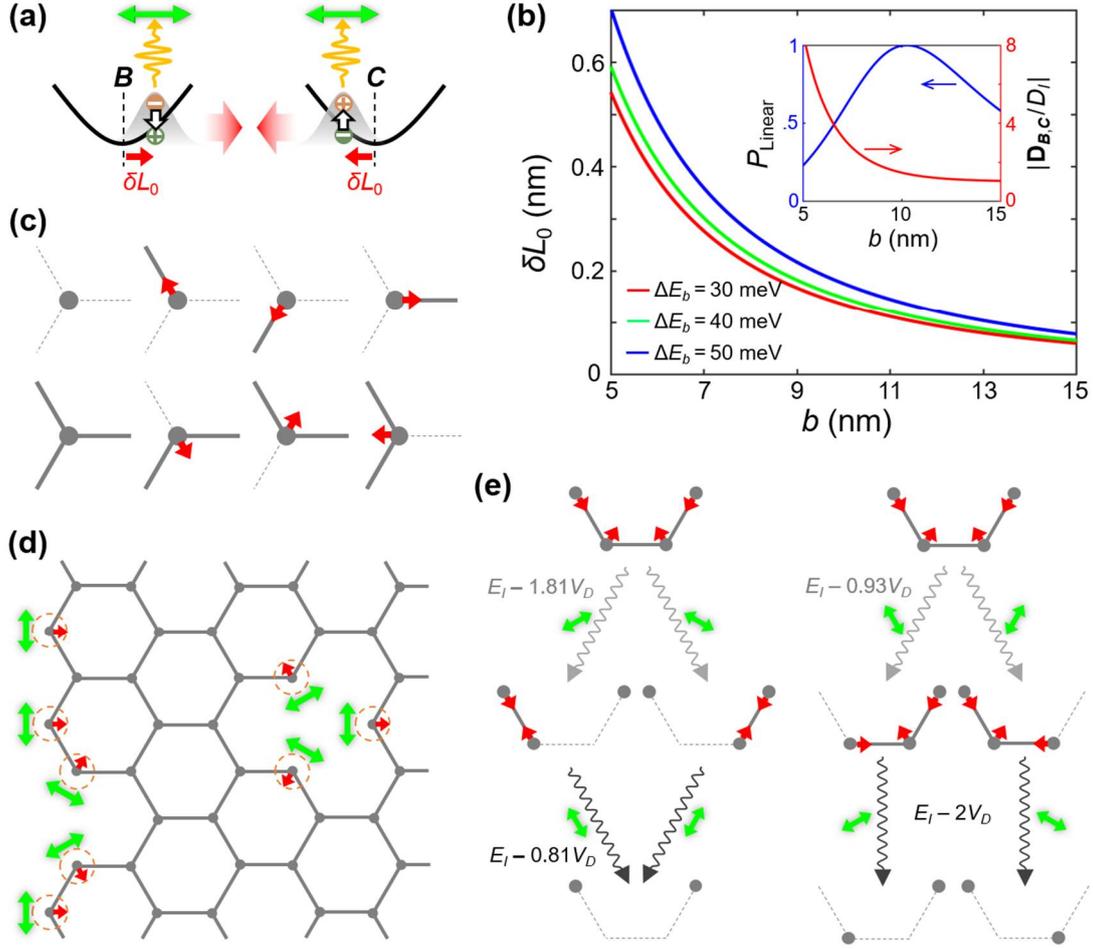

**Fig. 3 Anomalous luminescence in a honeycomb superlattice with nearest-neighbor attractions.** (a) A tiny displacement of a nearest-neighbor pair by their attraction changes the emission polarization from circular to linear. (b) The equilibrium displacement $\delta L_0$ in (a), in moiré of different period $b$. The three curves correspond to different values of the binding energy difference $\Delta E_b$. The inset shows the degree of linear polarization and optical dipole strength of the displaced exciton wavepacket. (c) The dipolar force experienced by an exciton is unbalanced when its three nearest-neighbor sites are partially occupied, leading to an equilibrium displacement depending on the filling configurations. Thick solid (thin dashed) lines denote the presence (absence) of the nearest-neighbor dipolar attractions. (d) At full filling, the unbalanced sites only appear at the edges and vacancies, where excitons have much faster emission with linear polarizations (green double arrows), in contrast to the balanced sites in the bulk where excitons are dark. (e) Cascaded emission of a lowest-energy 4-exciton cluster, which produces entangled photons. The polarization and energy of the photons are shown, with $E_I$ the energy of the trapped exciton.

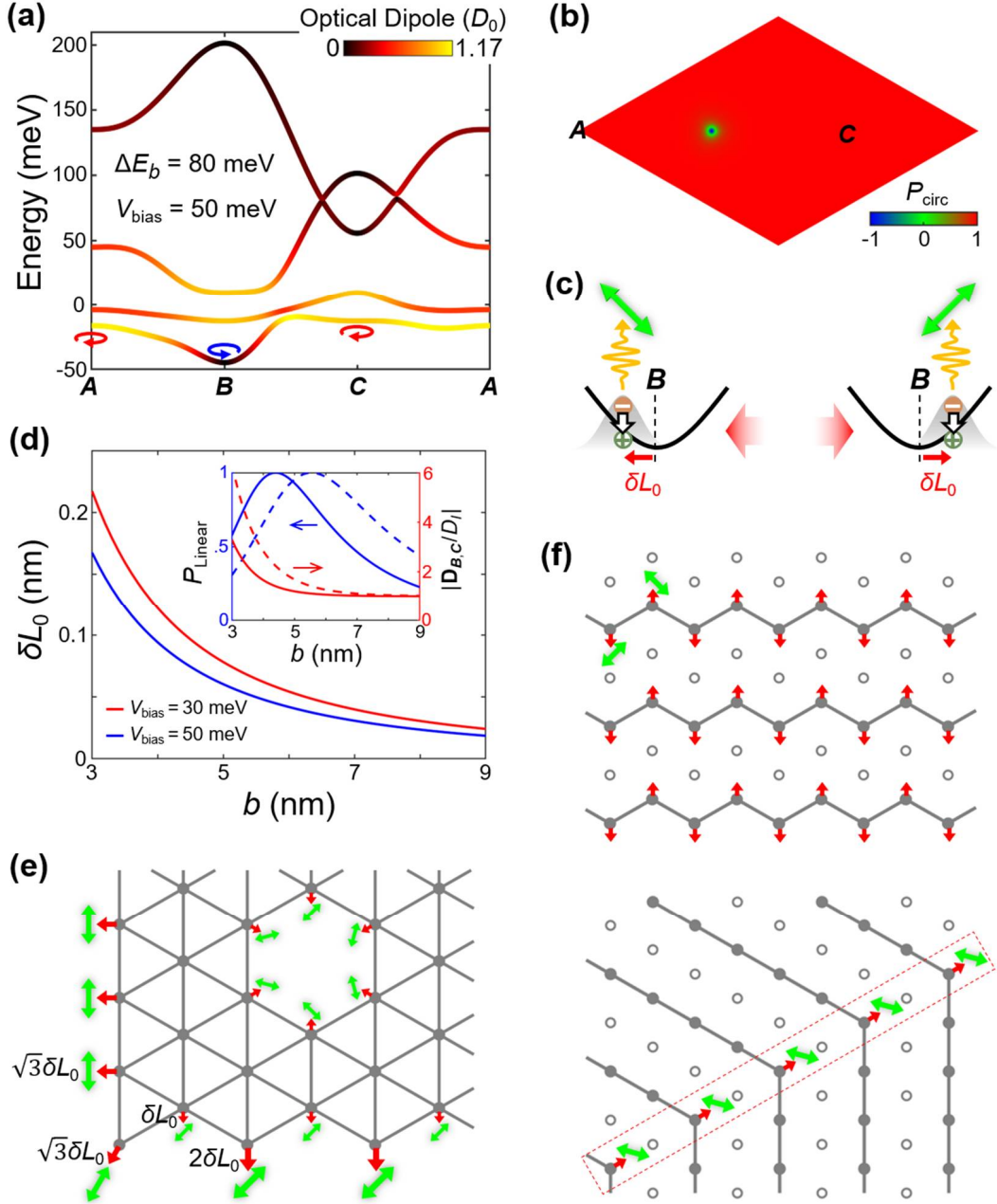

**Fig. 4 Anomalous luminescence in a triangular lattice of repulsive interactions.** (a) Moiré superlattice potentials for the four decoupled exciton branches under a larger binding energy difference $\Delta E_b = 80$ meV, and an interlayer bias $V_{bias} = 50$ meV that lifts the degeneracy at **B** and **C**. (b) Emission polarizations in the lowest energy branch, changing rapidly in the neighborhood of the trapping site **B**. (c) In this triangular lattice, the dipolar repulsion between nearest-neighbor **B** sites pushes the two trapped excitons away from each other by a tiny displacement, which results in orthogonal linear polarizations of their emissions. (d) The equilibrium $\delta L_0$ in moiré of various period $b$ under two $V_{bias}$ values. The inset shows the degree of linear polarization and optical dipole strength for a site-displacement of $\delta L_0$ (solid) and $2\delta L_0$ (dashed) under $V_{bias} = 50$ meV. (e) The displacements (red arrows) and emission linear polarizations (green double arrows) for the unbalanced sites at the edges and vacancies under full filling. The equilibrium displacement can be $\delta L_0$, $\sqrt{3}\delta L_0$ or $2\delta L_0$. (f) Two distinct stripe phases under the filling factor 1/2, where radiative recombination occurs in the bulk and at domain boundaries respectively. Solid (empty) dots denote the occupied (unoccupied) sites.

## Appendix A. Intralayer component of the trapped exciton wavepacket and its contribution to optical dipole: a symmetry analysis

Both X$_{intra}$ and X$_{inter}$ have Wannier type wavefunctions with the electron and hole localized around $\pm K$ in momentum-space, so here we use the envelope approximation to write the involved conduction (valence) band Bloch function as $\psi_{K+k,c}(r_e) \approx e^{ik \cdot r_e}\psi_{K,c}(r_e)$ ($\psi_{K'+k,v'}(r_e) \approx e^{ik \cdot r_h}\psi_{K',v'}(r_h)$). The X$_{inter}$ momentum eigenstate (also eigenstate of the electron-hole Coulomb interaction) is [39]

$$X_Q^{(inter)}(r_e, r_h) = \sum_{\Delta Q} \Phi_I(\Delta Q)\psi_{K+\frac{m_e}{M_0}Q+\Delta Q, c}(r_e)\psi^*_{K'-\frac{m_h}{M_0}Q+\Delta Q, v'}(r_h)$$
$$\approx e^{iQ \cdot R}\Phi_I(r_{eh})\psi_{K,c}(r_e)\psi^*_{K',v'}(r_h).$$

Here $R \equiv \frac{m_e}{M}r_e + \frac{m_h}{M}r_h$ is the center-of-mass (COM) coordinate of the exciton, and $r_{eh} \equiv r_e - r_h$ corresponds to the electron-hole relative coordinate. $\Phi_I(r_{eh}) \equiv \sum_{\Delta Q} \Phi_I(\Delta Q)e^{i\Delta Q \cdot r_{eh}}$ is the wavefunction of the electron-hole relative motion. Since the exciton has a large binding energy ~200 meV, the slowly modulating moiré potential barely affects $\Phi_I(r_{eh})$. So only the lowest energy electron-hole relative wavefunction (1s state) need to be considered.

An X$_{inter}$ wavepacket is then the linear superposition of the momentum eigenstates:

$$W_{R_c}^{(inter)}(r_e, r_h) = \sum_Q e^{-iQ \cdot R_c}W_n(Q)X_Q^{(inter)}(r_e, r_h)$$
$$\approx W_n(R - R_c)\Phi_I(r_{eh})\psi_{K,c}(r_e)\psi^*_{K',v'}(r_h).$$

Here $W_n(R - R_c) \equiv \sum_Q e^{iQ \cdot (R-R_c)}W_n(Q) \propto \frac{1}{w}e^{-\frac{(R-R_c)^2}{2w^2}}$ is the COM envelope in real space, with $n$ denoting the ground state ($s$-type), first excited state ($p$-type), etc., and $R_c$ denoting its center. $w$ is the half-width of the COM envelope.

Besides the slowly varying $W_n(R - R_c)$ and $\Phi_I(r_{eh})$ parts, the X$_{inter}$ wavepacket also contains the Bloch part $\psi_{K,c}(r_e)\psi^*_{K',v'}(r_h)$. Note that $W_n(R - R_c)$ and $\Phi_I(r_{eh})$ are eigenfunctions of $\hat{C}_3$-rotation about the wavepacket center $R_c$. In contrast, the $\psi_{K,c}(r_e)\psi^*_{K',v'}(r_h)$ part can be an eigenfunction of $\hat{C}_3$ only when the rotation center $R_c$ is at a high-symmetry point, i.e. $A$, $B$ or $C$ in the moiré (Fig. 1a), and the $\hat{C}_3$ eigenvalues are $+1$, $-1$ and $0$ respectively for $R_c$ being at $A$, $B$ and $C$ [31].

An X$_{intra}$ wavepacket, on the other hand, has a simpler structure:

$$W_{n,R_c}^{(intra)}(r_e, r_h) \approx W_n(R - R_c)\Phi(r_{eh})\psi_{K,c}(r_e)\psi^*_{K,v}(r_h).$$

With the electron and hole Bloch functions in the same layer, the entire wavefunction is always an eigenfunction under $\hat{C}_3$-rotation about the wavepacket center $R_c$ (regardless of whether $R_c$ is displaced from $B/C$ or not), and eigenvalues is determined by the index $n$ only (i.e. $s$-type or $p$-type COM envelope).

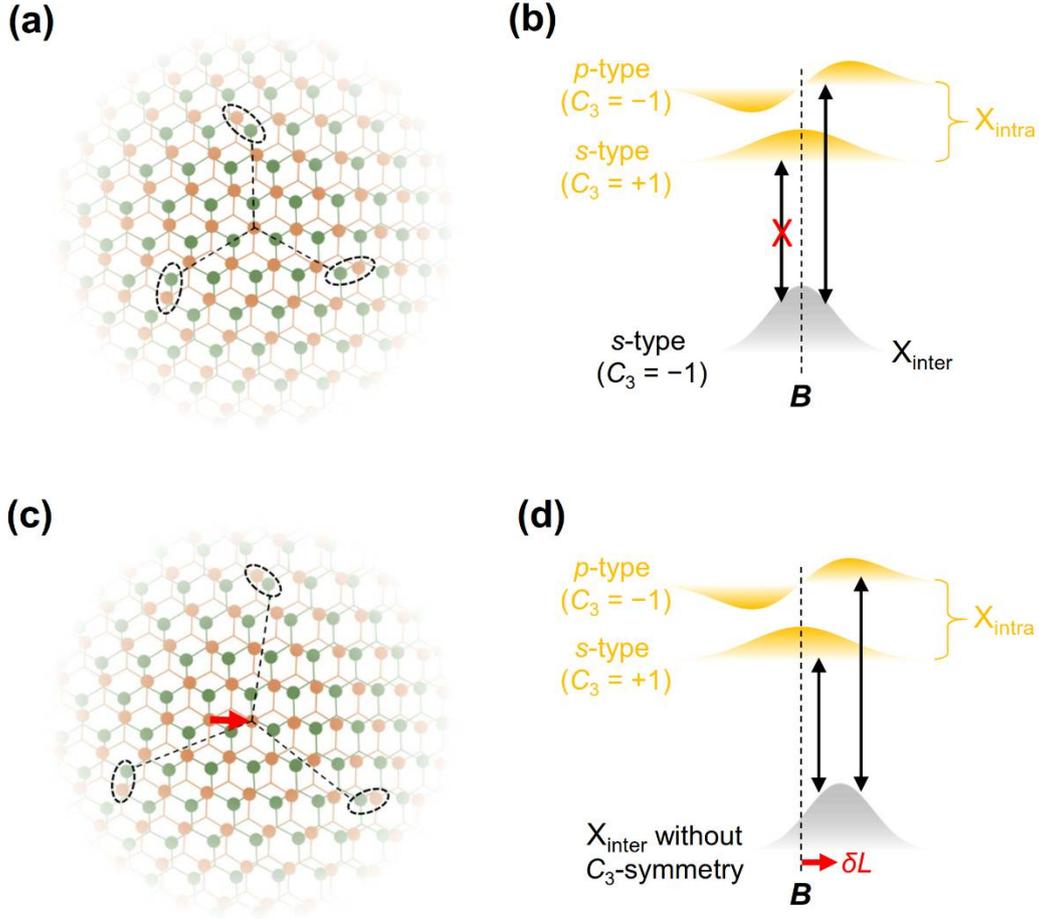

**Fig. 5** (a) The bilayer atomic registry underlying an exciton wavepacket centered at $B$, with a spread out determined by the COM envelope function as well as the relative motion. Each orange/green dot denotes a metal atom $d$-orbital in the upper/lower layer, which is the main constituent of the band edge Bloch function $\psi_{\mathbf{K},c}(\mathbf{r}_e)/\psi_{\mathbf{K'},v'}(\mathbf{r}_h)$. The three circled areas are related by $\hat{C}_3$-rotation, and the whole wavepacket is $\hat{C}_3$-symmetric. (b) The exciton wavepacket centered at $B$. The wavepacket is primarily an $X_{inter}$ with the $s$-type envelope. With the $\hat{C}_3$-symmetry of the trapping site, the wavefunction is an eigenfunction of $\hat{C}_3$, and its eigenvalue ($C_3 = -\mathbf{1}$) corresponds to the circular polarization ($\sigma-$) of the emission. Due to the perturbative interlayer coupling (vertical double arrows), the wavepacket also contains a $p$-type envelope $X_{intra}$ component with $C_3 = -\mathbf{1}$. $X_{intra}$ with an $s$-type envelop has $C_3 = +\mathbf{1}$ and emits with $\sigma+$ polarization instead, therefore is not allowed in the wavepacket by the $\hat{C}_3$-symmetry. (c) The atomic registry underlying an exciton wavepacket slightly displaced from $B$, with the red arrow denoting the displacement vector. The three circled areas are no longer related by $\hat{C}_3$-rotation, thus the wavepacket is not $\hat{C}_3$-symmetric. (d) The exciton wavepacket with a small displacement from $B$. As the restriction imposed by the $\hat{C}_3$-symmetry is lifted, both the $p$- and $s$-type envelope $X_{intra}$ components are allowed in the wavepacket wavefunction. In (b) and (d), grey color denotes the envelope function of the primary $X_{inter}$ component, and yellow color denotes that of $X_{intra}$. The latter can only have a small contribution in the ground state wavefunction as they are far detuned in energy.

Therefore for an $X_{inter}$ wavepacket with its center being at a high-symmetry point (**B**/**C** as concerned in this work), albeit the spread out in real space as determined by $W_n(\mathbf{R} - \mathbf{R}_c)$ and $\Phi_I(\mathbf{r}_{eh})$, the wavefunction does have the perfect $\hat{C}_3$-symmetry about its center $\mathbf{R}_c$, as can be seen in Fig. 5a. This $\hat{C}_3$-symmetry does not allow an *s*-type envelope $X_{inter}$ to be hybridized with an *s*-type envelope $X_{intra}$ that has a difference $\hat{C}_3$ eigenvalue. It can hybridize only with a *p*-type envelope $X_{intra}$ having the same $\hat{C}_3$ eigenvalue (see Fig. 5b). The *p*-type envelope $X_{intra}$ however, has a negligibly small optical dipole. As a result, the overall optical dipole of the exciton wavepacket remains circularly polarized, with a magnitude essentially being just that of the $X_{inter}$ optical dipole.

When slightly displaced from **B**/**C**, however, because of the incommensurability of the two layers, the Bloch part $\psi_{\mathbf{K},c}(\mathbf{r}_e)\psi^*_{\mathbf{K}',v'}(\mathbf{r}_h)$ is no longer an eigenfunction of $\hat{C}_3$-rotation about the displaced wavepacket center $\mathbf{R}_c$, so the overall $X_{inter}$ wavefunction no longer has the $\hat{C}_3$-symmetry (c.f. Fig. 5c). The broken $\hat{C}_3$-symmetry lifts the restriction on the coupling between $X_{inter}$ and $X_{intra}$, and the *s*-type envelope $X_{inter}$ wavepacket can also hybridize in a small portion of the *s*-type envelope $X_{intra}$ wavepacket (see Fig. 5d, and Supplementary Section II). It is this displacement-dependent coupling together with the large optical dipole of *s*-type envelope $X_{intra}$ that dramatically change the wavepacket optical properties, even if the displacement is much smaller than the spread of the wavefunction in real space.

**Appendix B. Interaction between two exciton wavepackets**

We consider two *s*-type envelope $X_{inter}$ wavepackets $W_B^{(inter)}(\mathbf{r}_{e1},\mathbf{r}_{h1})$ and $W_C^{(inter)}(\mathbf{r}_{e2},\mathbf{r}_{h2})$ located at a nearest-neighbor pair of **B** and **C** sites, respectively. They interact through the Coulomb potential

$$\hat{V}_{XX} = V_{inter}(\mathbf{r}_{ee}) + V_{inter}(\mathbf{r}_{hh}) - V_{intra}(\mathbf{r}_{e1h2}) - V_{intra}(\mathbf{r}_{e2h1}).$$

Here $V_{inter}(\mathbf{r}) = \frac{e^2}{4\pi\epsilon\sqrt{r^2+d^2}}$ and $V_{intra}(\mathbf{r}) = \frac{e^2}{4\pi\epsilon r}$ correspond to the interlayer and intralayer Coulomb interactions, respectively, see Fig. 6a. $\mathbf{r}_{ee} \equiv \mathbf{r}_{e1} - \mathbf{r}_{e2}$, $\mathbf{r}_{hh} \equiv \mathbf{r}_{h1} - \mathbf{r}_{h2}$, $\mathbf{r}_{e1h1} \equiv \mathbf{r}_{e1} - \mathbf{r}_{h1}$, $\mathbf{r}_{e2h2} \equiv \mathbf{r}_{e2} - \mathbf{r}_{h2}$, $\mathbf{r}_{e1h2} \equiv \mathbf{r}_{e1} - \mathbf{r}_{h2}$, $\mathbf{r}_{e2h1} \equiv \mathbf{r}_{e2} - \mathbf{r}_{h1}$. Generally $V_{inter}(\mathbf{r}) < V_{intra}(\mathbf{r})$ due to the vertical separation $d$ between the two layers.

It is known that the matrix elements of $\hat{V}_{XX}$ can be separated into the dipole-dipole and exchange interaction parts [23]. The dipole-dipole interaction part corresponds to the process $(e1,h1) + (e2,h2) \rightarrow (e1,h1) + (e2,h2)$, see Fig. 6b. Whereas The exchange interaction part corresponds to the process $(e1,h1) + (e2,h2) \rightarrow (e1,h2) + (e2,h1)$, see Fig. 6c. Below we derive their forms separately.

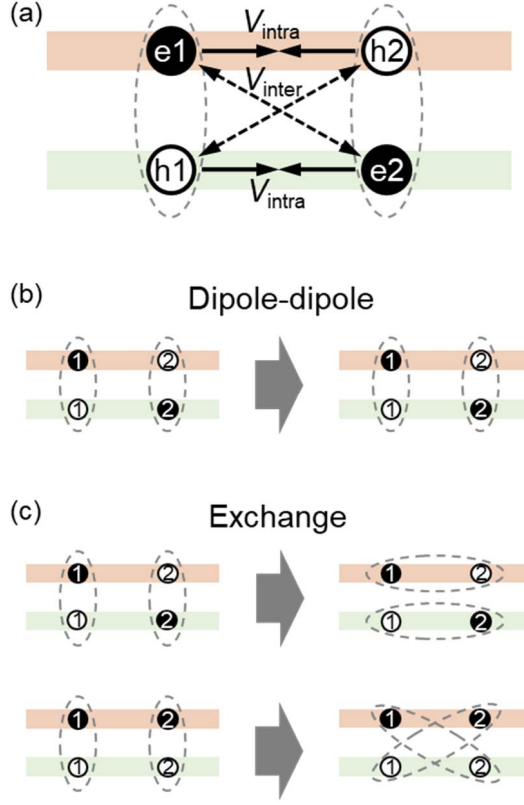

**Fig. 6** (a) The Coulomb interaction between two $X_{\text{inter}}$ $(e1, h1)$ and $(e2, h2)$. Solid/empty circles denote the electrons/holes. $V_{\text{inter}}$ and $V_{\text{intra}}$ correspond to the interlayer and intralayer Coulomb interactions, respectively. (b) Schematic illustration of the dipole-dipole interaction, which corresponds to the process $(e1, h1) + (e2, h2) \rightarrow (e1, h1) + (e2, h2)$. (c) Schematic illustration of the exchange interaction, which corresponds to the process $(e1, h1) + (e2, h2) \rightarrow (e1, h2) + (e2, h1)$. Note that different electric dipole orientations of $(e1, h1)$ and $(e2, h2)$ can result in $(e1, h2)$ and $(e2, h1)$ with either intralayer (upper row) or interlayer configurations (lower row).

### *i. The dipole-dipole interaction part*

The strength of the dipole-dipole interaction is

$$V_{\text{dd}} \equiv \left\langle W_B^{(\text{inter})}(\mathbf{r}_{e1}, \mathbf{r}_{h1}) W_C^{(\text{inter})}(\mathbf{r}_{e2}, \mathbf{r}_{h2}) \left| \hat{V}_{XX} \right| W_B^{(\text{inter})}(\mathbf{r}_{e1}, \mathbf{r}_{h1}) W_C^{(\text{inter})}(\mathbf{r}_{e2}, \mathbf{r}_{h2}) \right\rangle$$

$$= \int d\mathbf{r}_{e1} d\mathbf{r}_{e2} d\mathbf{r}_{h1} d\mathbf{r}_{h2} \left| \psi_{\mathbf{K},c}(\mathbf{r}_{e1}) \psi^*_{\mathbf{K}',v'}(\mathbf{r}_{h1}) \psi_{\mathbf{K},c}(\mathbf{r}_{e2}) \psi^*_{\mathbf{K}',v'}(\mathbf{r}_{h2}) \right|^2$$

$$\times \left| W(\mathbf{R}_1 - B) W(\mathbf{R}_2 - C) \Phi_I(\mathbf{r}_{e1h1}) \Phi_I(\mathbf{r}_{e2h2}) \right|^2$$

$$\times \left( V_{\text{inter}}(\mathbf{r}_{ee}) + V_{\text{inter}}(\mathbf{r}_{hh}) - V_{\text{intra}}(\mathbf{r}_{e1h2}) - V_{\text{intra}}(\mathbf{r}_{e2h1}) \right)$$

$$\approx \int d\mathbf{R}_1 d\mathbf{R}_2 d\mathbf{r}_{e1h1} d\mathbf{r}_{e2h2} \left| W(\mathbf{R}_1 - B) W(\mathbf{R}_2 - C) \Phi_I(\mathbf{r}_{e1h1}) \Phi_I(\mathbf{r}_{e2h2}) \right|^2$$

$$\times \left[ V_{\text{inter}} \left( \mathbf{R}_1 - \mathbf{R}_2 + \frac{m_h}{M}(\mathbf{r}_{e1h1} - \mathbf{r}_{e2h2}) \right) + V_{\text{inter}} \left( \mathbf{R}_1 - \mathbf{R}_2 - \frac{m_e}{M}(\mathbf{r}_{e1h1} - \mathbf{r}_{e2h2}) \right) \right.$$
$$- V_{\text{intra}} \left( \mathbf{R}_1 - \mathbf{R}_2 + \frac{m_h}{M} \mathbf{r}_{e1h1} + \frac{m_e}{M} \mathbf{r}_{e2h2} \right)$$
$$\left. - V_{\text{intra}} \left( \mathbf{R}_1 - \mathbf{R}_2 - \frac{m_e}{M} \mathbf{r}_{e1h1} - \frac{m_h}{M} \mathbf{r}_{e2h2} \right) \right].$$

Here $\mathbf{R}_1 \equiv \frac{m_e}{M}\mathbf{r}_{e1} + \frac{m_h}{M}\mathbf{r}_{h1}$ and $\mathbf{R}_2 \equiv \frac{m_e}{M}\mathbf{r}_{e2} + \frac{m_h}{M}\mathbf{r}_{h2}$. Since both the wavepacket half-width $w$ and the Bohr radius $a_B$ are much larger than the monolayer lattice constant, in the last step above we have approximated the fast oscillating terms $|\psi_{\mathbf{K},c}(\mathbf{r}_e)|^2$ and $|\psi^*_{\mathbf{K}',v'}(\mathbf{r}_h)|^2$ by their mean values $1 = \int |\psi_{\mathbf{K},c}(\mathbf{r}_e)|^2 d\mathbf{r}_e = \int |\psi^*_{\mathbf{K}',v'}(\mathbf{r}_h)|^2 d\mathbf{r}_h$.

Note that the two nearest-neighbor wavepacket centers are separated by $b/\sqrt{3} \gg w, a_B$. For an order of magnitude estimation we can consider the limit of $w \to 0$ and $a_B \to 0$, and write $|W(\mathbf{R})|^2 \approx \delta(\mathbf{R})$, $|\Phi_I(\mathbf{r}_{eh})|^2 \approx \delta(\mathbf{r}_{eh})$. Then the dipole-dipole interaction is of a simple form

$$V_{\mathrm{dd}} \approx 2V_{\mathrm{inter}}(B-C) - 2V_{\mathrm{intra}}(B-C)$$
$$= -\frac{e^2}{2\pi\epsilon}\left(\frac{1}{R_{BC}} - \frac{1}{\sqrt{R_{BC}^2 + d^2}}\right) = -\frac{e^2 d^2}{4\pi\epsilon}\left(\frac{b}{\sqrt{3}}\right)^{-3} + O\left(\left(\frac{d}{b}\right)^4\right).$$

Here $\epsilon$ is the average dielectric constant of the surrounding substrate.

The corrections from the small but finite $w$ and $a_B$ are analyzed by a series expansion of $V_{\mathrm{inter/intra}}(\mathbf{R}_{BC} + \mathbf{r})$ near $\mathbf{R}_{BC} \equiv B - C$,

$$V_{\mathrm{inter}}(\mathbf{R}_{BC} + \mathbf{r}) = \frac{e^2}{4\pi\epsilon\sqrt{R_{BC}^2 + d^2}}\left[1 - \frac{2\mathbf{R}_{BC}\cdot\mathbf{r} + r^2}{2(R_{BC}^2 + d^2)} + \frac{3}{2}\frac{(\mathbf{R}_{BC}\cdot\mathbf{r})^2}{(R_{BC}^2 + d^2)^2} + O\left(\left(\frac{r}{b}\right)^3\right)\right],$$

$$V_{\mathrm{intra}}(\mathbf{R}_{BC} + \mathbf{r}) = \frac{e^2}{4\pi\epsilon R_{BC}}\left[1 - \frac{2\mathbf{R}_{BC}\cdot\mathbf{r} + r^2}{2R_{BC}^2} + \frac{3}{2}\frac{(\mathbf{R}_{BC}\cdot\mathbf{r})^2}{R_{BC}^4} + O\left(\left(\frac{r}{b}\right)^3\right)\right].$$

Note that both $|W(\mathbf{R})|^2$ and $|\Phi_I(\mathbf{r}_{eh})|^2$ are even functions, thus all the odd terms of $\mathbf{R}_1 - B$, $\mathbf{R}_2 - C$, $\mathbf{r}_{e1h1}$ and $\mathbf{r}_{e2h2}$ vanish after integration. The dipole-dipole interaction up to the 2nd order of $\frac{w}{b}$ and $\frac{a_B}{b}$ can then be derived as

$$V_{\mathrm{dd}} \approx -\frac{e^2 d^2}{4\pi\epsilon}\left(\frac{b}{\sqrt{3}}\right)^{-3}\left(1 + \frac{27}{2}\frac{w^2}{b^2} + \frac{81}{8}\frac{m_e^2 + m_h^2}{M^2}\frac{a_B^2}{b^2}\right).$$

On the other hand, the trapped exciton near $B/C$ contains a small component of $X_{\mathrm{intra}}$, which reduces its electric dipole. Using the perturbative wave function Eq. (4), we find the electric dipole is reduced by a fraction

$$\int d\delta\mathbf{L}|W(\delta\mathbf{L})|^2\left[\left(\frac{J_v}{2\Delta_v - \Delta E_b}\frac{\delta L}{b}\right)^2 + \left(\frac{J_c}{2\Delta_c - \Delta E_b}\frac{\delta L}{b}\right)^2\right] = \left[\left(\frac{J_v}{2\Delta_v - \Delta E_b}\right)^2 + \left(\frac{J_c}{2\Delta_c - \Delta E_b}\right)^2\right]\frac{w^2}{b^2}.$$

After taking into account this contribution, the dipole-dipole interaction up to the 2nd order of $\frac{w}{b}$ and $\frac{a_B}{b}$ is

$$V_{\mathrm{dd}} \approx -\frac{e^2 d^2}{4\pi\epsilon}\left(\frac{b}{\sqrt{3}}\right)^{-3}\left(1 + \left(\frac{27}{2} - 2\left(\frac{J_v}{2\Delta_v - \Delta E_b}\right)^2 - 2\left(\frac{J_c}{2\Delta_c - \Delta E_b}\right)^2\right)\frac{w^2}{b^2} + \frac{81}{8}\frac{m_e^2 + m_h^2}{M^2}\frac{a_B^2}{b^2}\right).$$

The wavepacket width scales as $w \propto \sqrt{b}$ ($w \sim 1.4$ nm when $b = 8$ nm from our estimation), whereas $a_B \sim 1$-$2$ nm is nearly independent of $b$. So $\frac{a_B^2}{b^2}$ increases faster than

$\frac{w^2}{b^2}$ when decreasing $b$. Furthermore, the coefficient of the $\frac{w^2}{b^2}$ term is much smaller than that of the $\frac{a_B^2}{b^2}$ term ($\left(\frac{J_v}{2\Delta_v - \Delta E_b}\right)^2 + \left(\frac{J_c}{2\Delta_c - \Delta E_b}\right)^2 \sim 7$ from our estimated parameters). Therefore, for a relatively small moiré period, we expect the correction from $\frac{a_B^2}{b^2}$ dominates over that from $\frac{w^2}{b^2}$, resulting in an overall enhancement to the interaction strength.

### *ii. The exchange interaction part*

The strength of the exchange interaction is

$$V_{\text{ex}} \equiv \frac{1}{2} \left\langle W_B^{(\text{inter})}(\mathbf{r}_{e1}, \mathbf{r}_{h2}) W_C^{(\text{inter})}(\mathbf{r}_{e2}, \mathbf{r}_{h1}) \left| \hat{V}_{XX} \right| W_B^{(\text{inter})}(\mathbf{r}_{e1}, \mathbf{r}_{h1}) W_C^{(\text{inter})}(\mathbf{r}_{e2}, \mathbf{r}_{h2}) \right\rangle$$

$$+ \frac{1}{2} \left\langle W_B^{(\text{inter})}(\mathbf{r}_{e2}, \mathbf{r}_{h1}) W_C^{(\text{inter})}(\mathbf{r}_{e1}, \mathbf{r}_{h2}) \left| \hat{V}_{XX} \right| W_B^{(\text{inter})}(\mathbf{r}_{e1}, \mathbf{r}_{h1}) W_C^{(\text{inter})}(\mathbf{r}_{e2}, \mathbf{r}_{h2}) \right\rangle$$

$$\approx \int d\mathbf{r}_{e1} d\mathbf{r}_{e2} d\mathbf{r}_{h1} d\mathbf{r}_{h2}\, \Phi_I(\mathbf{r}_{e1h1}) \Phi_I^*(\mathbf{r}_{e1h2}) \Phi_I(\mathbf{r}_{e2h2}) \Phi_I^*(\mathbf{r}_{e2h1})$$

$$\times W(\mathbf{R}_1 - \mathbf{B}) W(\mathbf{R}_2 - \mathbf{C}) W\left(\frac{m_e}{M_0}\mathbf{r}_{e1} + \frac{m_h}{M_0}\mathbf{r}_{h2} - \mathbf{B}\right) W\left(\frac{m_e}{M_0}\mathbf{r}_{e2} + \frac{m_h}{M_0}\mathbf{r}_{h1} - \mathbf{C}\right)$$

$$\times \left(V_{\text{inter}}(\mathbf{r}_{ee}) + V_{\text{inter}}(\mathbf{r}_{hh}) - V_{\text{intra}}(\mathbf{r}_{e1h2}) - V_{\text{intra}}(\mathbf{r}_{e2h1})\right).$$

For $b/\sqrt{3} \gg w, a_B$, $V_{\text{ex}} \approx 4 \left|\Phi_I(\mathbf{R}_{BC}) W\left(\frac{m_h}{M_0}\mathbf{R}_{BC}\right)\right|^2 \left(V_{\text{inter}}(\mathbf{R}_{BC}) - V_{\text{intra}}(\mathbf{R}_{BC})\right)$. Since $|\Phi_I(\mathbf{R}_{BC})|^2 \ll 1$ and $\left|W\left(\frac{m_h}{M_0}\mathbf{R}_{BC}\right)\right|^2 \ll 1$, $V_{\text{ex}}$ is thus exponentially small compared to $V_{\text{dd}}$. Furthermore, X$_{\text{inter}}$ trapped at $B$ and $C$ have opposite electric dipole orientations, thus $(e1, h2)$ and $(e2, h1)$ correspond to X$_{\text{intra}}$ (see Fig. 6c), and the energy detuning further suppresses this off-diagonal process by exchange. Thus the exchange interaction between two exciton wavepackets located at $B$ and $C$ can be safely ignored.

On the other hand, for two excitons in the same trapping site, both the dipole-dipole and exchange parts can be important and dependent on $w$ and $a_B$ of the trapped exciton wavepacket [23]. However, such configuration is energetically unfavorable due to the strong onsite repulsion. So when the moiré superlattice is loaded with excitons at filling factor $\le 1$, one can just focus on the many-body configurations with exciton per trapping site less or equal to one, where the exciton-exciton interaction is dominated by the dipole-dipole interaction $V_{\text{dd}}$.

### Appendix C. Relaxation process and radiative emission of the trapped excitons

Consider a ground state exciton wavepacket trapped at $B$. When all its three nearest-neighbors are occupied, the dipolar force is balanced and the exciton wavepacket center $\mathbf{R}_c$ is right at $B$ with zero displacement. The ground state in the trap $|g\rangle$ is primarily an X$_{\text{inter}}$ with *s*-type COM envelope $W_g(\mathbf{R} - \mathbf{R}_c)$ (see Fig. 7a), and the first excited state $|e\rangle$ is primarily an X$_{\text{inter}}$ with a *p*-type envelope $W_e(\mathbf{R} - \mathbf{R}_c)$. We consider the evolution of this exciton wavepacket triggered by the annihilation of one of its nearest-neighbors, which makes the dipolar force unbalanced. The unbalanced dipole force changes the potential profile experienced by the exciton wavepacket, shifting the energy minimum to a new location slightly displaced from $\boldsymbol{B}$ by $\delta L$ (c.f. Fig. 7b). In this shifted potential trap, the new ground state $|g'\rangle$ and first excited state $|e'\rangle$ can be perturbatively expressed in terms of $|g\rangle$ and $|e\rangle$: $|g'\rangle \equiv |g\rangle - \delta_F |e\rangle$ and $|e'\rangle \equiv |e\rangle + \delta_F |g\rangle$. Note that $|g'\rangle$ and $|e'\rangle$ now describe exciton wavepackets whose centers are displaced from $\boldsymbol{B}$ by $\delta L \propto \delta_F$, see Fig. 7b. Other higher energy excited states in the new potential are ignored for simplicity, while including them do not change the picture.

Denoting the time for the annihilation event of its nearest-neighbor as $t = 0$. For $t \leq 0$ the exciton wavepacket has the wavefunction $|g\rangle$, which, for $t > 0$ is no longer an energy eigenstate, but rather the superposition of the new ground and first excited states: $|g\rangle = |g'\rangle + \delta_F |e'\rangle$ (see Fig. 7b below). A major mechanism responsible for the relaxation from $|e'\rangle$ to $|g'\rangle$ (Fig. 7c) is by phonon scattering. At low temperature $k_B T \ll \hbar\omega \sim 10$ meV, the latter being the quantization energy in the moiré trap, the phonon emission rate $\tau^{-1}$ can be obtained through the Fermi golden rule

$$\tau^{-1} = \frac{2\pi}{\hbar} \sum_{\alpha,\mathbf{q}} |C_{\alpha,\mathbf{q}}|^2 \delta(\hbar\omega - \hbar\omega_{\alpha,\mathbf{q}}).$$

Here $C_{\alpha,\mathbf{q}} \equiv \langle g' \otimes \mathbf{1}_{\alpha,\mathbf{q}} | \hat{H}_{X-\text{phonon}} | e' \otimes \mathbf{0} \rangle$ is the exciton-phonon coupling matrix element, where $|e' \otimes \mathbf{0}\rangle$ is the exciton excited state in the moiré trap plus the phonon vacuum and $|g' \otimes \mathbf{1}_{\alpha,\mathbf{q}}\rangle$ is the exciton ground state $|g'\rangle$ plus a $\alpha$-mode phonon with wave vector $\mathbf{q}$ ($\alpha$ can be LA/TA phonon in the upper or lower layer). $\omega_{\alpha,\mathbf{q}}$ corresponds to the phonon dispersion.

Using the electron-phonon coupling strength from early works [73,74], we estimate that the relaxation from $|e'\rangle$ to $|g'\rangle$ happens in a timescale of picosecond (for a detailed derivation, see Supplementary Section III). Such a timescale is orders faster as compared to the radiative decay of the exciton wavepacket centered either at $\boldsymbol{B}$ (where radiative decay takes hundreds of ns) or the displaced location (where radiative decay takes ~10 ns). So the subsequence radiative emission will predominately come from $|g'\rangle$, i.e., the ground state wavepacket with the displaced center (c.f. Fig. 7d).

The same picture given above also applies if the exciton starts as a ground state wavepacket slightly displaced from $\boldsymbol{B}$ (as in an initially unbalanced site). Again, the key is the relaxation to the new excitonic ground state (upon the sudden change of the dipolar force from nearest-neighbors) occurs on a timescale orders faster as compared to the radiative decay of the exciton wavepacket, the latter still takes ~10 ns for wavepacket centered at the displaced location.

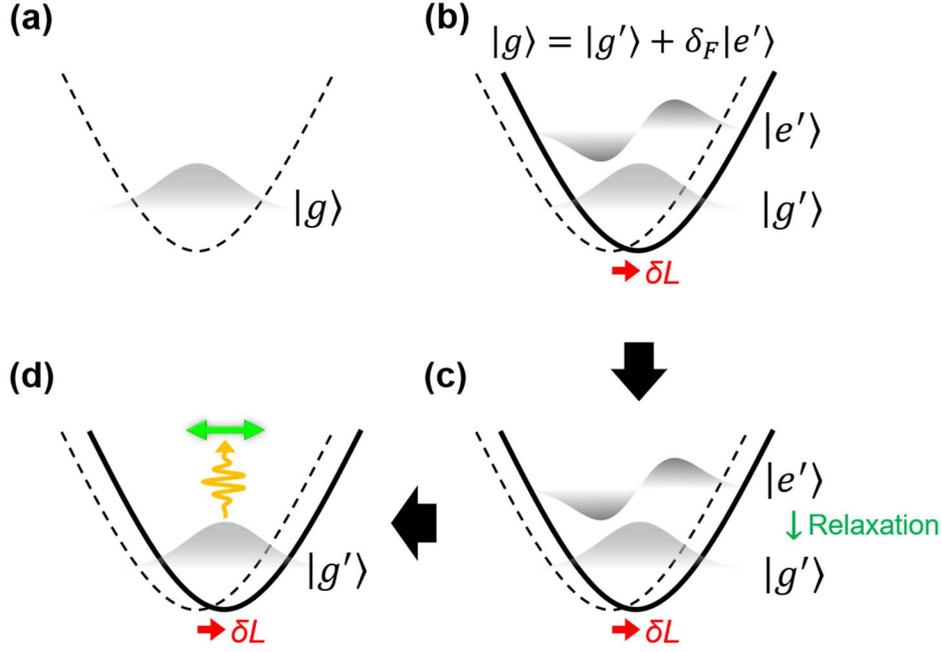

**Fig. 7** (a) A ground state exciton wavepacket initially at a balanced site, where the potential profile for the COM motion is denoted by the dashed curve, centered at ***B***. (b) After the annihilation of a nearest-neighbor, the exciton concerned feels an unbalanced dipolar force which changes the potential profile to the solid curve with a displaced center. The ground state and first excited state in this new potential profile are denoted as $|g'\rangle$ and $|e'\rangle$, respectively. The initial wavefunction is their superposition: $|g\rangle = |g'\rangle + \delta_F |e'\rangle$. (c) Through phonon scattering, the $|e'\rangle$ constituent of the wavefunction relaxes to the ground state $|g'\rangle$ in a time scale orders faster than the radiative decay of either $|g'\rangle$ or $|e'\rangle$. (d) The subsequence radiative emission from $|g'\rangle$. Note that the envelop functions shown in this figure are all for the primary $X_{inter}$ component of the exciton wavepacket.

    Concerning the radiative emission from the excitonic ground state in the moiré trap, we note that both the energy detuning and off-diagonal coupling between $X_{intra}$ and $X_{inter}$ are much larger than the broadening induced by their interaction with either the phonon or photon bath. The low energy excitonic eigenstates (e.g. $|g\rangle$ and $|e\rangle$) are spectrally well separated from each other by $\hbar\omega \sim 10$ meV, each having a definite form of superposition between the primary $X_{inter}$ component and a small $X_{intra}$ component. Experiments have found that the moiré trapped excitons in heterobilayers have spectral widths ~ 0.1 meV [34-36]. In this regime, the approach of diagonalizing the Hermitian Hamiltonian (Eq. (1)) and adding decay perturbatively is justified. Energy relaxation occurs between these spectrally well-separated excitonic eigenstates (c.f. Fig. 7c-d), and the radiative emission of the excitonic ground state is described by a single decay rate determined by its wavefunction, in which the form of the superposition between the primary $X_{inter}$ and the small $X_{intra}$ components dictates the optical dipole.

**Appendix D. Possible experimental schemes for isolating a small cluster of excitons**

The atomically thin geometry of the bilayer and the dipolar nature of the exciton point to several possibilities to isolate a small cluster of excitons trapped in a few moiré supercells, as schematically illustrated in Fig. 8:

(1) Using a biased metal tip (c.f. Fig. 8a), or local gate [60], to create a low energy region of a few moiré periods for isolating a cluster of dipolar excitons.
(2) Creating a strained region, by placing the bilayer on engineered substrate (c.f. Fig. 8b). Tensile strain lowers the energies of both $X_{intra}$ and $X_{inter}$ [61], so the strained region can act as a large scale potential trap that contains a few moiré supercells. In Ref. [26], such locally strained regions are engineered by placing heterobilayer on nanopillars (c.f. Fig. 8b), which can trap $X_{inter}$ with a tunable number from 1 to 5.
(3) Engineering the surrounding dielectric environment (c.f. Fig. 8c). For example, a $\sim O(10)$ nm scale graphene coverage can be realized by electron beam lithography to pattern a mask, and followed by oxygen plasma etching to etch the unwanted graphene. Ref. [62] shows that covering TMDs by a bilayer graphene (with a large dielectric constant $\varepsilon_2$) can significantly reduce the band gap as well as the exciton energy. The few moiré supercells underneath the graphene coverage are then energetically favored by the dipolar excitons, which can trap an exciton cluster.

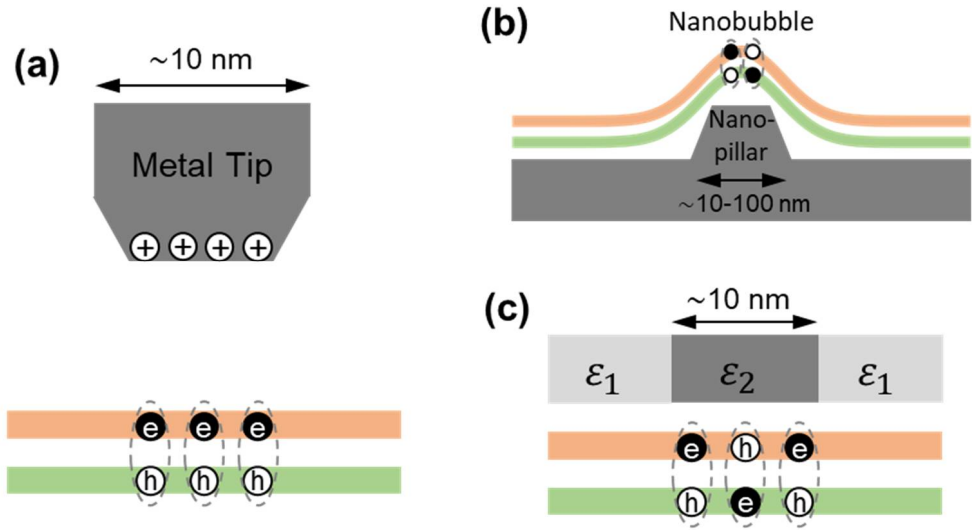

**Fig. 8** (a) A positively (negatively) charged nanoscale metal tip can trap the dipolar excitons at ***B*** sites with electric dipole down (***C*** sites with electric dipole up). Alternatively, the role of the movable tip can be played by a local gate. (b) The local tensile strain in a nanobubble can trap both ***B***- and ***C***-site excitons, which can be engineered by placing the bilayer on a nanopillar. (c) By engineering the local dielectric environment, the excitons can be trapped to the area with a larger dielectric constant.